\documentclass[epj,referee,a4paper]{svjour}
\usepackage{bm}
\usepackage[T1]{fontenc}
\usepackage[latin1]{inputenc}
\usepackage{graphicx}
\usepackage{amssymb}
\usepackage{esint}

\begin{document}

\title{Analytical study of non-linear transport across a semiconductor-metal junction}
\subtitle{Resonances, surface states, and non-linear transport}
\author{N. M. R. Peres}

\institute{ Department of Physics and Center of Physics, University of
Minho, PT-4710-057, Braga, Portugal, \email{peres@fisica.uminho.pt}}

\abstract{
In this paper we study analytically a one-dimensional model for a semiconductor-metal
junction. We study the formation of Tamm states and how they evolve when the semi-infinite
semiconductor and metal are coupled together. The non-linear current, as a function of the
bias voltage, is studied using the non-equilibrium Green's function method and the
density matrix of the interface is given. The electronic occupation of the sites
defining the interface has strong non-linearities as  function of the bias voltage due
to strong resonances present in the Green's functions of the junction sites.
The surface Green's function is computed 
analytically by solving
a quadratic matrix equation, which does not require adding a small imaginary constant to the
energy. The wave function for the surface states is given.
}


\maketitle

\section{Introduction}
The electronic properties of surfaces and interfaces has many fascinating features,
associated with the formation of surface states, modification of the band structure
and corresponding density of states, and electronic transport \cite{Ibach}.
Strongly localized surface states are known since a long time to be present at the
interface of semiconductors and metals \cite{Tamm} and have been shown to
influence the conductance of
 scanning tunneling microscopy (a quasi-one-dimensional process) \cite{Kobayashi}.
Also these states have recently been found in one-dimensional metamagnetic materials
\cite{Kivshar}.
Therefore the characterization of these states is an important aspect of the physics of surfaces of materials
and interfaces between two different solids.

In the fields of one-dimensional physics, specially those related to device applications,
the transport properties are of crucial importante, and therefore it is important to study
the effect that surfaces and interfaces have on these properties. Of particular interest
to us is the non-linear transport through a semiconductor-metal interface. The non-linear
transport requires the use of non-equilibrium methods, which were first introduced in this
context by Caroli {\it et al.} \cite{Caroli}.   Following these early developments, the
calculation of the current through a two-band system, connected to one-dimensional metals,
including the effect of disorder on the semiconductor was soon performed \cite{Glick},
and latter revisited by other authors \cite{Bishop}. As old as these investigations
may be, transport across one-dimensional metal-semiconductor-metal
systems is still an active  research topic \cite{Mathur},
specially in the field of quasi one-dimensional organic conductors.

Except for very simple models, such as transport across an impurity
\cite{Sautet88,Mizes91,Khomyakov05,Sols}, the transport calculations using the
non-equilibrium Green's functions method are all numerical. This is so due
to the fact that both the surface Green's function of the contacts and the Green's
functions
of the system (often called device) are obtained by the inversion of very
large matrices, a process that in general has no analytical solution.
There is even the wide-spread idea that the only surface Green's function that
can be computed analytically is that of a semi-infinite one-dimensional chain \cite{Sankey} (the surface Green's functions
of a cubic lattice reduce to the solution of a onedimensional problem with analytical solution \cite{Ferry}). 

Usually, the surface Green's function of a given lead is computed using a recursive
method developed long ago \cite{Rubio}. 
The convergence of this method is constrainted by the value of a
small imaginary positive number added to the energy of the Green's function.  There is however
an alternative method to compute the surface Green's function that does not depend
on the value of that small imaginary part; it even works with the small imaginary 
part equal to zero, a limit that is implicit in the definition of the Green's function
\cite{Dy,Dy2}.
As we show bellow, this method can be used to obtain analytical expressions for the
surface Green's function of a system with two atoms per unit cell, by solving a
quadratic matrix equation.

It is our purpose, in this paper, to study analytically, within
a simple toy model, 
the formation of Tamm states at the interface of a semiconductor-metal junction and its
non-linear transport, as a function of the bias voltage across such interface.
We certainly recognize our study to be that of a toy model problem, but the fact
that all quantities can be computed analytically makes our results relevant for
the study of more sophisticated models where no analytical solution exists. 
We will compute all the needed Green's function analytically, which in turn
give analytical expressions for the tunneling probability and for the system density matrix.
To our best knowledge this study has not been done before. 

The paper is organized as follows, in Sec. \ref{formalism} we formulate the problem in terms
of an infinite dimensional block tridiagonal matrix and explain how to derive the needed Green's
function as well as the method of solution for the surface Green's functions. 
In Sec. \ref{equilibrium} we study the existence of surface states in the semiconductor and
metal surfaces and how these states evolve to sharp resonances when the semiconductor and the
metal are coupled. In Sec. \ref{transport} we compute the charge current through the 
interface for arbitrarily large bias voltage and the local electron occupation at the
sites of the interface. Finally, in Sec. \ref{conclusions} we give our conclusions.
\section{ Hamiltonian toy model and formalism}
\label{formalism}

\begin{figure}[!hbpt]
\includegraphics[width=9cm]{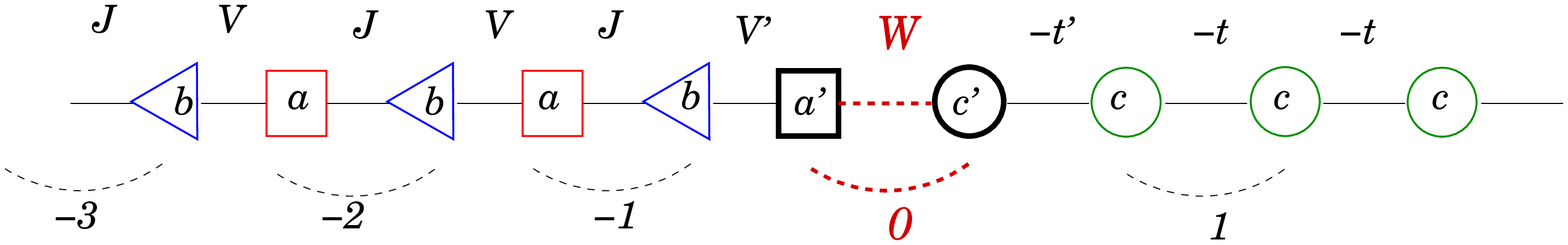}
\caption{(color online) Model of a semiconductor-metal interface.
The parameters we used through the text are:
$\epsilon_c=0.1$, $J=0.8$, $V=0.7$, $\epsilon_a=0.1$, $\epsilon_b=0.2$,
$t'=0.9$, $V'=0.9V$, $\epsilon_a'=1.1\epsilon_a$, $\epsilon_c'=5.0\epsilon_c$,
where all energies are in units of $t$. The parameter $W$ will be varied, but is also
given in units of $t$.} 
\label{fig:hamiltonian}
\end{figure}

In order to give a  analytical description of the  properties of a semiconductor-metal
junction we consider here a one-dimensional model. In Fig. \ref{fig:hamiltonian} we depict
the model used. The unit cells have two atoms. In the case of the semiconductor
the atoms are different ($a$ and $b$ in Fig. \ref{fig:hamiltonian}); in the metal
all the atoms are equal ($c$ in Fig. \ref{fig:hamiltonian}). The doubling of the unit
cell in the metal is only a matter of convenience, since it allow us to write the
Hamiltonian as a tridiagonal block matrix (see below). The junction takes place in the
unit cell $n=0$. The terminal atoms of the semiconductor and the metal are represented
by the letters $a'$ and $c'$, respectively. In the semiconductor the hopping
within a unit cell is $J$ and between unit cells is $V$. In the metal the hopping is $-t$.
The local energies in the semiconductor are $\epsilon_a$ and $\epsilon_b$ for atoms
$a$ and $b$, respectively, and for the metal atoms 
the local energy is represented by $\epsilon_c$. In general we expect that the atoms at the
surface of a metal, say, have different local energies and be connected to the bulk atoms by different
hopping parameters. This fact introduces the parameters $\epsilon_a'$, $\epsilon_c'$,
$V'$, $-t'$. The hopping across the interface is $W$. The reader may note that the
model has a somewhat large number of parameters. This is necessary, since at the
interface of both the metal and the semiconductor both the hopping and the
on-site energies are modified relatively to their bulk values. Naturally, the number
of parameters can be greatly reduced if we do not take into consideration the
effect of the surface on the model parameters. Although our study is aimed
to a general description of the effects of surfaces on the non-linear transport
across an interface, its application  to a real system is a  possibility; this would
 require us to use
the values of the parameters appropriate to the system under study. We note,
however, that all energies are given in units of $t$, and therefore the parameter
values used in the simulations can apply to a general system.

If we denote the semiconductor Hamiltonian by $H_L$, the metal Hamiltonian by $H_R$, and
the junction Hamiltonian by $H_d$ (defined by unit cell $n=0$), 
the full Hamiltonian of the problem, written in the
local atomic basis, reads
\begin{equation}
 H=
\left[
\begin{array}{ccc}
H_L & V_L & 0\\
V^\dag_L & H_d & V^\dag_R\\
0 & V_R & H_R
\end{array}
\right]
\label{hamiltonian33}
\end{equation}
where the matrices $V_L$ and $V_R$ are the coupling between the semiconductor and the
junction and the metal and junction, respectively. If we write $H_L$ and $H_R$ explicitly,
the Hamiltonian (\ref{hamiltonian33}) acquires a block tridiagonal form and reads 

\begin{equation}
 H=
\left[
\begin{array}{ccccccc}
\ddots & & & & & &\\
C^\dag_L & H_{sm} & C_L & & & &\\
 & C^\dag_L & H_{sm} & \Lambda_L & & &\\
 & & \Lambda^\dag_L & H_d & \Lambda_R & &\\ 
&  & & \Lambda^\dag_R & H_m & C_R & \\
& &  & & C^\dag_R & H_m & C_R \\
 &  & &  &  & & \ddots 
\end{array}
\right]
\label{h_tridiag}
\end{equation}
where the several matrices are given by
\begin{equation}
 H_{sm}=
\left[
\begin{array}{cc}
\epsilon_a & J \\
J & \epsilon_b \\
\end{array}
\right]\,,\hspace{0.5cm}
 H_d=
\left[
\begin{array}{cc}
\epsilon_a' & W \\
W & \epsilon_c' \\
\end{array}
\right]\,,
\end{equation}

\begin{equation}
 H_m=
\left[
\begin{array}{cc}
\epsilon_c & -t \\
-t & \epsilon_c \\
\end{array}
\right]\,,\hspace{0.5cm}
 C_L=
\left[
\begin{array}{cc}
0 & 0 \\
V & 0 \\
\end{array}
\right]\,,
\end{equation}

\begin{equation}
 \Lambda_L=
\left[
\begin{array}{cc}
0  & 0 \\
V' & 0 \\
\end{array}
\right]\,,\hspace{0.5cm}
 \Lambda_R=
\left[
\begin{array}{cc}
0 & 0 \\
-t' & 0 \\
\end{array}
\right]\,,
\end{equation}
and
\begin{equation}
 C_R=
\left[
\begin{array}{cc}
 0  & 0 \\
-t & 0 \\
\end{array}
\right]\,.
\end{equation}
Since we want to study the properties of the junction, we need to compute local quantities.
This is best accomplished using Green's functions. The full Green's function of the system
is defined by
\begin{equation}
(\bm 1 E+i0^+-H)G^+=\bm 1\,, 
\end{equation}
where we have chosen the retarded function (denoted with the $+$
superscript), and $\bm 1$ is an infinite identity matrix. 
The matrix form of the Green's function is  
\begin{equation}
G^+= \left[
\begin{array}{ccc}
 G_{LL}  & G_{Ld} & G_{LR}\\ 
G_{dL}  & G_{dd} & G_{dR}\\
G_{RL}  & G_{Rd} & G_{RR}
\end{array}
\right]\,.
\end{equation}
The quantity of interest is $G_{dd}$, which is shown to have the form 
\begin{equation}
G_{dd}^+= (\bm 1 E+i0^+-H_d-\Sigma_L^+-\Sigma_R^+)^{-1}\,, 
\end{equation}
where the matrices $\Sigma_L^+$ and $\Sigma_R^+$ are the self energies and have the form
\begin{equation}
 \Sigma_L^+=\Lambda^\dag_L G^+_{LL}\Lambda_L\,,
\hspace{0.5cm}
\Sigma_R^+=\Lambda_R G^+_{RR}\Lambda_R^\dag\,,
\end{equation}
where the Green's functions $G^+_{LL}$ and $G^+_{RR}$ are the surface Green's function
of the Hamiltonians $H_L$ and $H_R$, respectively. These Green's functions are defined as
\begin{equation}
\left[
\begin{array}{ccc}
\ddots& C_L & \\
 C^\dag_L& U_{sm} & C_L\\
 & C^\dag_L & U_{sm}
\end{array}
\right] 
\left[
\begin{array}{ccc}
\ddots& \vdots & \vdots \\
\ldots &G_{-2,-2} & G_{-2,L}\\
 \ldots& G_{L,-2} & G_{LL}
\end{array}
\right] =\bm 1\,,
\end{equation}
with $U_{sm}=E+i0^+-H_{sm}$ and a similar equation defining $G^+_{RR}$.
It is possible to find a close form for $G^+_{LL}$ and $G^+_{RR}$ \cite{Dy,Sankey},
reading
\begin{eqnarray}
\label{GLL}
G_{LL}^+ &=& [E+i0^+-H_{sm}-C^\dag_LG_{LL}C_L]^{-1}\,,\\
\label{GRR}
G_{RR}^+ &=& [E+i0^+-H_m-C_RG_{RR}C_R^\dag]^{-1}\,. 
\end{eqnarray}
(In the Appendix we give a simple derivation of Eqs. (\ref{GLL}) and (\ref{GRR}).)
The solution of (\ref{GLL}) and (\ref{GRR}) can in general be done
numerically only, by using a decimation procedure \cite{Rubio}, or a direct iterative
solution \cite{Sankey}. In these two methods it is necessary to introduce a small
imaginary part, that is $0^+$ is replaced by $\eta^+$, where $\eta^+$ is a finite
number. The rate of convergence of the two methods depend of the value of $\eta$,
which we would like to be as small as possible. There is however another method available which is
based on the solution of a quadratic matrix equation \cite{Dy,Dy2,Kim} and that does not
require the use of a finite value of $\eta$. This method was recently used in the
context of transport through molecular junctions \cite{Dahnovskya,Dahnovskyb}, but has been
essentially forgotten by the community working on surface Green's functions applied to non-linear transport.
Let us explain how this last method works considering, for this purpose,  the solution of 
$G_{LL}^+$. We start by defining an auxiliary quantity $Y=G_{LL}^+C_L$, which allows to write
Eq. (\ref{GLL}) as
\begin{equation}
\label{Y}
 C^\dag_LY^2+(H_{sc}-E-i0^+)Y+C_L=0\,. 
\end{equation}
We now assume that there is a similarity transformation $Q$ that diagonalizes the matrix $Y$,
defined as $Y=Q\bar YQ^{-1}$, where $\bar Y$ is a diagonal $2\times2$ matrix,
$\bar Y=\mathrm{ diag }(y_1,y_2)$. Using $\bar Y$
we can write Eq. (\ref{Y}) as
\begin{equation}
\label{Ya}
 C^\dag_LQ\bar Y^2+(H_{sc}-E-i0^+)Q\bar Y+C_LQ=0\,. 
\end{equation}
Writing $Q$ as $Q=(q_1,q_2)$, where $q_1$ and $q_2$
are vector columns of two elements, we obtain that
the solution for $\bar Y$ and $Q$ reduces to the solution of a quadratic eigenvalue 
problem of the form
\begin{equation}
\label{Yb}
 [C^\dag_Ly_i^2+(H_{sc}-E-i0^+)y_i+C_L]q_i=0\,,\hspace{0.5cm} i=1,2\,.  
\end{equation}
Equation (\ref{Yb}) has non-trivial solutions if the following determinant is zero
\begin{equation}
\label{eigen}
 \vert\vert C^\dag_Ly_i^2+(H_{sc}-E-i0^+)y_i+C_L\vert\vert=0\,.
\end{equation}
The solution of Eq. (\ref{eigen}) produces in principle four eigenvalues $y_i$, of
which only two are physical. 
In general the solution of the quadratic matrix equation has to done numerically, but 
for our $2\times2$ matrix an analytical solution exists. For real eigenvalues $y_i$,
convergence of the solution for $G_{LL}^+$ requires that $ y_i<1$. All the imaginary eigenvalues 
satisfy the condition $\vert y_i\vert =1$, that is the imaginary $y_i$'s can be written as
\begin{equation}
 y_i= x \pm i\sqrt{1-x^2}\,,\hspace{0.5cm}x^2<1\,,
\end{equation}
and the choice of the sign is made in order to satisfy the analytical
properties of $G_{LL}^+$, namely $\Im G_{LL}^+<0$. Alternatively, a small
positive imaginary part can be added to the energy, and the correct choice of $y_i$
are those solutions that lie in the unit circle.

\section{Local electronic properties at equilibrium}
\label{equilibrium}
In this section we want to address the electronic properties of the junction
when the chemical potential of the semiconductor and the metal are equal, and therefore
there is no current flowing through the system. To that end we need to compute
$G^+_{LL}$, $G^+_{RR}$, and $G^+_{dd}$; they are all 2$\times$2 matrices.
For the case of $G^+_{LL}$,  Eq. (\ref{eigen}) has the form
\begin{equation}
 y^2_i(E-\epsilon_a)(E-\epsilon_b)-y_i(Jy_i+V)(J+Vy_i)=0\,,
\end{equation}
which has three solutions, $y_i=y_1=0$ and
\begin{equation}
y_i=\beta\pm\sqrt{\beta^2-1}\,, 
\end{equation}
with $\beta=[(E-\epsilon_a)(E-\epsilon_b)-J^2-V^2]/(2VJ)$. For $\beta>1$ the correct choice
of $y_i$ is 
\begin{equation}
 y_i=y_2=\beta-{\mathrm sgn}(\beta)\sqrt{\beta^2-1}\,,
\end{equation}
and for $\beta<1$, the correct choice is
\begin{equation}
y_i=y_2=\beta-i\sqrt{1-\beta^2}\,. 
\end{equation}
For $y_1$ the eigenvector $q_1$ is
\begin{equation}
 q_1=\left[
\begin{array}{c}
0\\
v_1            
\end{array}
\right]
\end{equation}
where $v_1$ in any real number. For $y_2$, the eigenvector $q_2$
is 
\begin{equation}
 q_2=\left[
\begin{array}{c}
u_2\\
v_2            
\end{array}
\right]\,,
\end{equation}
with $u_2=v_2X$, and $X$ given by
\begin{equation}
X=\frac {J+y_2V}{E-\epsilon_a}=\frac{y_2(E-\epsilon_b)}{Jy_2+V}\,, 
\end{equation}
with $v_2$ any real number. It is now a simple task to compute $Q$ and its inverse, from
which $Y$, and therefore $G^+_{LL}$, is obtained. The surface Green's function is 
given by 
\begin{equation}
\label{GLLf}
 G^+_{LL}=\frac {y_2}{JV}
\left[\begin{array}{cc}
      E-\epsilon_b & J\\
	J & E-\epsilon_a-Vy_2/X 
      \end{array}
\right]\,.
\end{equation}
Using exactly the same procedure we obtain for $G^+_{RR}$ the equation
\begin{equation}
 \label{GRRf}
G^+_{RR}=\frac {y_2}{t^2}
\left[\begin{array}{cc}
      E-\epsilon_c +y_2tZ& -t\\
	-t & E-\epsilon_c 
      \end{array}
\right]\,,
\end{equation}
with
\begin{equation}
Z=\frac{t(1+y_2)}{y_2(\epsilon_c-E)}=\frac{\epsilon_c-E}{t(1+y_2)}\,, 
\end{equation}
and
\begin{equation}
y_2=\left\{
\begin{array}{c}
 \alpha-{\mathrm sgn}(\alpha)\sqrt{\alpha^2-1}\,,\hspace{0.5cm}\alpha^2>1\,,\\
\alpha-i\sqrt{1-\alpha^2}\,,\hspace{0.5cm}\alpha^2<1\,,
\end{array}
 \right.
\end{equation}
with $\alpha=[(E-\epsilon_c)^2-2t^2]/(2t^2)$. 

The calculation of $G^+_{dd}$ requires the determination of the self energies. These
are simply obtained as
\begin{equation}
\label{SLL}
\Sigma_L^+= 
\left[\begin{array}{cc}
      V'^2(G^+_{LL})_{22}& 0\\
	0 & 0
      \end{array}
\right]\,,
\end{equation}
and
\begin{equation}
\label{SRR}
\Sigma_R^+= 
\left[\begin{array}{cc}
     0& 0\\
  	0 &  t'^2(G^+_{RR})_{11}
      \end{array}
\right]\,.
\end{equation}
The matrix elements $(G^+_{LL})_{22}$ and $(G^+_{RR})_{11}$ are, after
some algebra, simply given
by
\begin{eqnarray}
\label{GLL2}
 (G^+_{LL})_{22}&=&(E-\epsilon_b)^{-1}(1+Jy_2/V)\,,\\
\label{GRR1} 
(G^+_{RR})_{11}&=&(E-\epsilon_c)^{-1}(1+y_2)\,.
\end{eqnarray}
At first sight, Eq. (\ref{GRR1}) does not look like the surface Green's function
of a semi-infinite one-dimensional chain (that is because we used two atoms per unit cell for the metal) \cite{peres}, however simple algebraic
manipulations show that $(G^+_{RR})_{11}$ can be put in the known form
\begin{equation}
 (G^+_{RR})_{11}=\frac {E-\epsilon_c}{2t^2}-\frac {i}{2t^2}\sqrt{4t^2-(E-\epsilon_c)^2}\,,
\end{equation}
for $(E-\epsilon_c)^2<4t^2$ and a similar equation for $(E-\epsilon_c)^2>4t^2$.
Using Eqs. (\ref{SLL}) and (\ref{SRR}), $G^+_{dd}$ is given by 
\begin{equation}
\label{Gddf}
G_{dd}^+=\frac {1}{S(E)} 
\left[\begin{array}{cc}
     S_c(E)& W\\
  	W & S_a(E)
      \end{array}
\right]\,,
\end{equation}
with $S(E)=S_a(E)S_c(E)-W^2$, and
\begin{eqnarray}
S_a(E)&=&E-\epsilon_a'- V'^2(G^+_{LL})_{22}\,,\\
S_c(E)&=&E-\epsilon_c'- t'^2(G^+_{RR})_{11}\,.
\end{eqnarray}
If $W=0$, the two systems are decoupled, and $(G_{dd}^+)_{11}$ is the surface
Green's function of the semiconductor and  $(G_{dd}^+)_{22}$ is the surface
Green's function of the metal. 

In some conditions, the existence of surfaces in a material give rise to surface states.
In semiconductors these states lie in the 
gap of the semiconductor. These type of states are determined by
the condition $S(E)=0$, with $S(E)$ a real number. When $S(E)$
is imaginary, $S(E)=S_1(E)+iS_2(E)$, resonant states are determined
from $S_1(E)=0$; $S_2(E)$ will be related to the width of the resonance.
The local density of states at the junction atoms is given by (either with zero or finite $W$)
\begin{eqnarray}
\rho_{a'}=-\frac 1{\pi}\Im (G^+_{dd})_{11}\,,\\
\rho_{c'}=-\frac 1{\pi}\Im (G^+_{dd})_{22}\,,
\end{eqnarray}
where $\rho_{a'}$ and $\rho_{c'}$ are the local density of states at $a'$ and
$c'$ atoms, respectively. 

\begin{figure}[!hbpt]
\includegraphics[width=7cm]{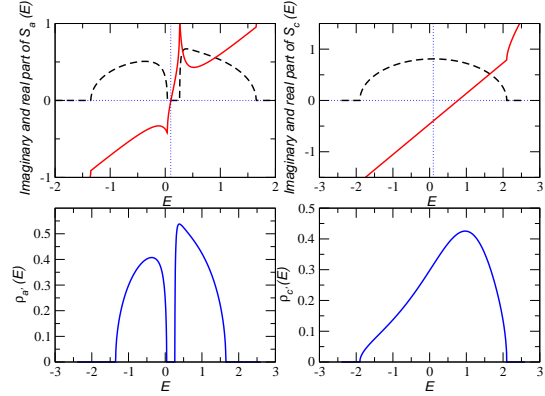}
 \caption{\label{fig:resonanceW0}(color online) Plot of the imaginary (dashed line)
and the real parts (solid line) of $S_a(E)$ and $S_c(E)$. To
make the resonance in the metal well visible we have used $\epsilon_c'=5\epsilon_c$.
In the lower panels we have the local density of states at atoms $a'$ and $c'$ when $W=0$.
The other parameters are those introduced in Fig. \ref{fig:hamiltonian}.}

\end{figure}

Considering first the case $W=0$, the case in which the two systems are decoupled, the
surface states of the semiconductor satisfy the condition $S_a(E)=0$, with
$S_a(E)$ real, and the
resonances satisfy the condition $\Re S_a(E)=0$. Similar expression hold for the metal with
$S_a(E)$ replaced by $S_c(E)$. In Fig. \ref{fig:resonanceW0} we plot the
imaginary and the real parts of $S_a(E)$ and $S_c(E)$.

In the case of the metal, the condition $\Re S_c(E)=0$ corresponds to the maximum
of the density of states, which represents a very broad resonance. In the case of
the semiconductor we see that in the gap we have $S_a(E)=0$, with
$S_a(E)$ real, thus corresponding to a pole in the Green's function 
and, therefore, representing a surface or Tamm state.
Let us calculate the wave function of the Tamm state (for simplicity
we consider the case $\epsilon_{a'}=0$). The general wave function has the
form
\begin{equation}
 \vert\psi\rangle = \sum_{m=-\infty}^{-1}(a_m\vert m,a\rangle +b_m\vert m,b\rangle )\,,
\end{equation}
where $\vert m,a\rangle$ and $\vert m,b\rangle$ are position-base states
at the sites of the chain, and $a_m$ and $b_m$ are the corresponding amplitudes.
The matrix elements of the Hamiltonian, $\langle n,a \vert H \vert\psi\rangle$
and $\langle n,b \vert H \vert\psi\rangle$ leads to
\begin{eqnarray}
 V'b_{-1}\delta_{-1,n}+Vb_n\theta(-n-1)
+Jb_{n+1}=Ea_n\,,\\
Ja_{n-1}+Va_n=Eb_n\,,
\end{eqnarray}
where $\theta(x)$ is the Heavyside function, with $\theta(0)=0$.
The above equations are subject to the boundary condition
$b_0=0$.
We now make the observation that for $E=0$, there is a non trivial
solution for the amplitudes $a_m$ given by the recursive relation
\begin{equation}
 Ja_{m-1}+Va_{m}=0\,,
\end{equation}
and all $b_m=0$. The wave function of the surface state is therefore
given by
\begin{equation}
 \vert\psi\rangle = \sum_{m=-\infty}^{-1}a_{1}(-V/J)^{\vert m\vert-1}\vert m,a\rangle\,,
\end{equation}
with $a_1=\sqrt{1-(V/J)^2}$ obtained from the normalization of the wave function.
Clearly, for the surface state to exist we need $V/J<1$, the case we used
in our numerical calculations. For $V/J\ge 1$ the surface state is absent.
In the particular case $V/J> 1$ a Shockley state will develop in the gap,
which is not a surface state. Also for the metal there are some conditions
where localized states can form at the surface. Let us take the semi-infinite
one-dimensional metal introduced in Fig. \ref{fig:hamiltonian}, whose Hamiltonian
reads
\begin{equation}
 H=-t\sum_{n=0}^{\infty}(\vert n><n+1\vert + H.c.)+
\vert 0><0\vert\epsilon_{c'}\,,
\end{equation}
and we have assumed all the hoppings equal. Proposing a localized
wave function of the form
\begin{equation}
 \psi_{\rm loc}=A\sum_{n=0}^{\infty}e^{-n\lambda}\vert  n>\,,
\end{equation}
and writing the Schr\"odinger equation as
\begin{eqnarray}
-ta_1&=&(E-\epsilon_{c'})a_0\,,\\
Ea_n&=&-t(a_{n-1}+a_{n+1}),\hspace{0.5cm} n>0\,, 
\end{eqnarray}
we obtain for the energy of the localized state
\begin{equation}
 E=-2t\cosh\lambda\,,
\end{equation}
with $\lambda$ the solution of
\begin{equation}
 e^{\lambda}=\epsilon_{c'}/t\,.
\end{equation}
Since $\lambda$ must be larger than zero we must have $\epsilon_{c'}/t>1$ and the energy
of the localized state is 
\begin{equation}
E=-2\frac{\epsilon_{c'}^2+t^2}{\epsilon_{c'}}\,, 
\end{equation}
located below the bottom of the metal band. So, in this special condition
it is possible for the metal to develop a localized state at the surface.
If we generalize the above case and include the possibility that the hopping
between the site $c'$ and the site $c$ is $t'$, the wave function of the
localized state has to be generalized to

\begin{equation}
 \psi_{\rm loc}=A\sum_{n=1}^{\infty}e^{-n\lambda}\vert  n> + Ab\vert0>\,,
\end{equation}
and the Schr\"odinger equation has now the form
\begin{eqnarray}
-t'a_1&=&(E-\epsilon_{c'})a_0\,,\\
Ea_1&=&-t'a_0+ta_1\,,\\
Ea_n&=&-t(a_{n-1}+a_{n+1}),\hspace{0.5cm} n>1\,, 
\end{eqnarray}
whose solution gives $b=t/t'$ and
\begin{equation}
 2e^{-\lambda}=
\frac{t\epsilon_{c'}}{t'^2-t^2}\pm
\frac{1}{t'^2-t^2}
\sqrt{(t\epsilon_{c'})^2+4t^2(t'^2-t^2)}\,,
\end{equation}
and the energy is still given by $E=-2t\cosh\lambda$. Since we must have
$\lambda>0$ only some values of the parameters produce surface states,
in particular we must have $t'>t$ (which is not the case considered in the
simulations.).

\begin{figure}[!hbpt]
\includegraphics[width=7cm]{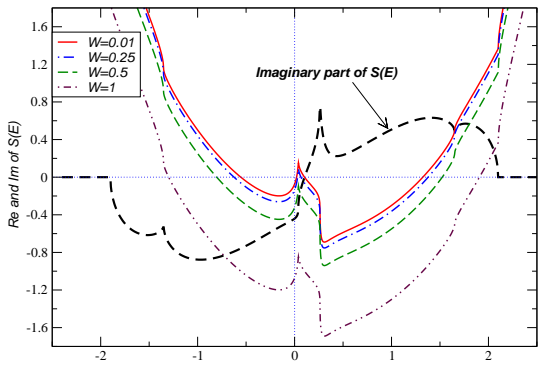}
 \caption{\label{fig:resonanceW}(color online) Plot of the imaginary 
and  real part of $S(E)$ for different values of $W$. 
The parameters are those introduced in Fig. \ref{fig:hamiltonian}.}
\end{figure}

\begin{figure}[!hbpt]
\includegraphics[width=7cm]{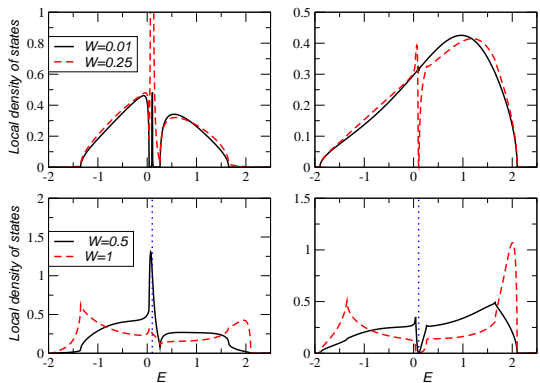}
 \caption{\label{fig:ldos} (color online) Plot of local density of states $\rho_{a'}$ and
$\rho_{c'}$ for different values of $W$. The left panels depict the semiconductor density of states
at site $a'$ and the right ones the same quantity for the metal at site
$c'$. At site $c'$ the formation of an anti-bound state is clear, where a dip in the
density of states is seen. The parameters are those introduced in Fig. \ref{fig:hamiltonian}. }
\end{figure}

We now make $W$ finite coupling the semiconductor and the metal.
In Fig. \ref{fig:resonanceW} we plot the
imaginary and the real parts of $S(E)$ for different values of $W$.
Because $S(E)$ contains now contributions from both the imaginary parts of $S_a(E)$
and $S_c(E)$ there will be a finite imaginary part  in the energy range of the
semiconductor's gap. As a consequence the Tamm state previously
located in the gap becomes now a sharp resonance and is visible in the local density of states
$\rho_{a'}$, as we show in Fig. \ref{fig:ldos}.
Of particular interest is the strong transfer of spectral weight from the density of states
of the metal to the resonance in the semiconductor, given rise to an anti-resonance in the
density of states of the metal. It is interesting that the high density of states persists
where the resonance was located even when $\Re S(E)$ is no longer zero, albeit small.
Only for large values of $W$, such as $W=1$, a large transfer of spectral weight 
to resonances at lower and higher energies takes place. From Fig. \ref{fig:resonanceW}
we see that for moderate values of $W$, even when we have $\Re S(E)=0$, no sharp resonance
appear in the local density of states because $\Im S(E)$ is very large at those points.
For $W=1$ strong resonances appear at the lower and top edges of the bands.

\section{Transport properties}
\label{transport}

We now study the non-equilibrium transport across the junction. 
This is done using the non-equilibrium 
Green's function method \cite{keldysh}. This method is particularly suited to study the regime where
the system has a strong departure from equilibrium, such as when the bias potential
$V_b$ is large. The system is however in the steady state.  Since the seminal paper of Caroli
{\it et al.} on non-equilibrium quantum transport \cite{Caroli}, that the
method of non-equilibrium Green's functions become generalized to the
calculation of transport quantities of nanostructures.
There are many places where one can find
a description of the method \cite{Ferry,Datta,Jauho}, 
but an elegant one was recently introduced
in the context of transport through systems that have bound states, showing that
the problem can be reduced to the solution of a kind of quantum Langevin equation \cite{Sen}.

The general idea of this method is that two perfect leads are coupled to our system,
which is usually called the device. In our case the device is defined by the junction,
a two site system, involving the $a'$ and $c'$ sites. The Green's function of the
device has to be computed in the presence of the leads. This corresponds to our $G_{dd}^+$ Green's
function. Besides the Green's function we need the effective coupling between the leads
and the system (the junction)
which are determined in terms of the self-energies as
\begin{equation}
\Gamma_{L/R} = \frac {i}{2\pi}(\Sigma^+_{L/R}-\Sigma^-_{L/R})\,, 
\end{equation}
therefore the effective coupling $\Gamma_{L/R}$ depends on the surface Green's
function of the perfect leads. According to the general theory the two leads are in thermal
equilibrium at temperatures $T_{L/R}$ and chemical potential $\mu_{L/R}$ and are connected
to the system at some time $t_0$. The bottom line is that the total current through the
device is given by (both spins included)
\begin{equation}
J=\frac {2e}{h}\int_{-\infty}^\infty dE T(E)[f(E,\mu_L,T_L)-f(E,\mu_R,T_R)]\,, 
\end{equation}
where the transmission $T(E)$ is given by 
\begin{equation}
 T(E)= 4\pi^2{\rm Tr}[\Gamma_LG^+_{dd}\Gamma_RG^-_{dd}]\,.
\end{equation}
Performing the trace we obtain
\begin{equation}
T(E)=4V'^2t'^2\Im (G_{LL}^+)_{22}\Im (G_{RR}^+)_{11}\frac {W^2}{\vert S(E)\vert^2}\,. 
\end{equation}
\begin{figure}[!hbpt]
\includegraphics[width=7cm]{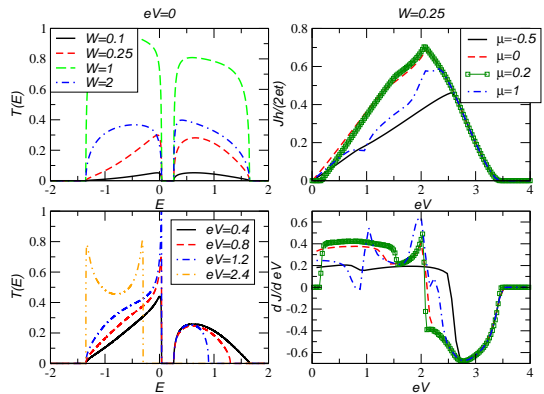}
 \caption{\label{fig:TE} (color online) Transmission $T(E)$ for $V=0$ and different values of $W$
(top left); $T(E)$ for different values of $eV$ and $W=0.25$ (down left).
Non-linear current $J$ as function of the bias energy $eV$ (top right), for different values
of the chemical potential and $W=0.25$. Differential conductance (bottom right), $d J/d\, (eV)$, 
for the same curves in top right panel; the negative values of the differential conductance
are due to the non-linearities of the current curves upon the bias potential.
The parameters are those introduced in Fig. \ref{fig:hamiltonian}.}
\end{figure}
Because we are applying a bias voltage across the junction, we choose
$\mu_L=\mu$ and $\mu_R=\mu-eV$. Also the electrostatic potential has to change
continuously between the semiconductor and the metal. We choose that the variation
of the potential is proportional to the distance to the electrodes, therefore the
local energies in the junction have to be modified according to
\begin{equation}
\epsilon_{a'}\rightarrow \epsilon_{a'} -eV/3\,,\hspace{0.5cm}
\epsilon_{c'}\rightarrow \epsilon_{c'} -2eV/3\,.
\label{EV}
\end{equation}
The sites of the metal are shift by $-eV$.
Since $T(E)$ is computed using $G_{dd}^+$ and this depends on $\epsilon_{a'}$
and on $\epsilon_{c'}$, $T(E)$ will also be a function of the bias potential
$V$. The choice made in Eq. (\ref{EV}) corresponds to the solution of the discrete
Poisson equation ignoring the charge fluctuations taking place to screen the
external electric field.
(In general terms, to determine the potential
at sites $a'$ and $c'$ we would have to solve the Poisson equation coupled to the
solution of the Schr\"odinger equation \cite{Hasegawa,Yonemitsu}, 
but the above transformation of
$\epsilon_{a'}$ and $\epsilon_{c'}$ gives a good first
approach to the exact result.)
One technical aspect worth stressing here is the fact that, in general, non-linear
transport should be done in a self-consistent way. This is the case in two situations:
(i) when interactions (Coulomb or phonons, say) are taken into account; (ii) when
the potential inside the conductor is relevant for transport. In the case we are
considering here the conductor is reduced to a two-site system in the absence
of interactions. Therefore,  a self-consistent calculation is not needed.

Since the transmission function $T(E)$ depends on the imaginary part of the surface density
of states, the gap due to semiconductor shows up. From Fig. \ref{fig:TE} we see that the
current $J$ across the junction is only linear upon $V$ for very small values of the bias
potential. As $V$ is further increased non-linearities in $J$ start to develop  due to
the
energy dependence of $T(E)$,
 which is strongly influenced by the resonances 
shifting in energy as we vary the bias (see left down panel
of Fig. \ref{fig:TE}); for large $V$ (say $\sim$ 2) the upper band of the
semiconductor non longer contributes to the transport, due to the relative
shift of the local density of states induced by $V$. Also at large $V$
 resonances develop in $T(E)$ at both lower and higher energies;
 this effect is seen at the bottom left panel of 
Fig. \ref{fig:TE} for $V=2.4$ eV. As the junction is strongly biased the current
starts being suppressed due to the relative motion of the local energy  of sites in the junction and the
bands of the semiconductor. For large bias, the on-site energy of the junction's sites
 moves to lower
energies and these states are no longer available
to the electrons coming from the semi-conductor.
A qualitative model for tunneling across a one-site system also presents the general
features seen for $J$ in Fig. \ref{fig:TE}, except that the effects due to resonances
 are not present \cite{Jauho}. 
Naturally, when the chemical potential of the semiconductor lies on the gap it is necessary
a finite bias voltage to produce a current. 

\begin{figure}[!hbpt]
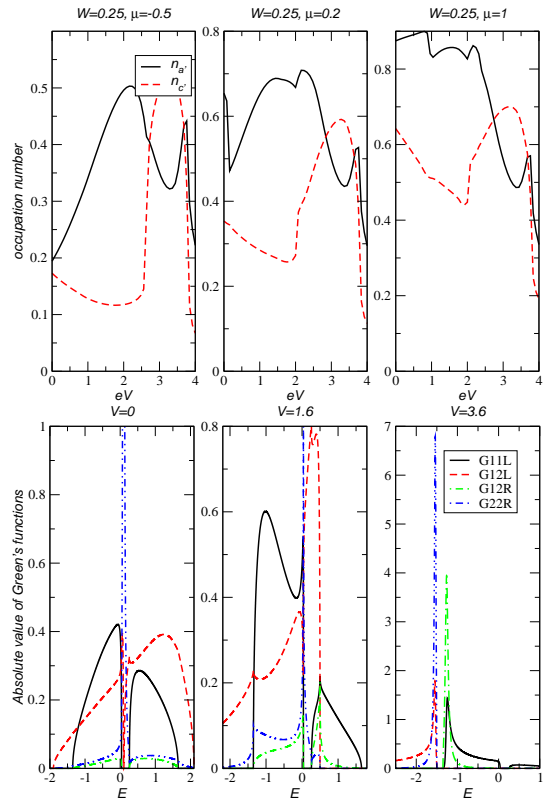

\includegraphics[width=7cm]{Fig_occupation.eps}
\includegraphics*[width=7cm]{Fig_G_matrix.eps}
 \caption{\label{fig:density}(color online) Top figure: occupation numbers $n_{a'}$ and $n_{c'}$ as function of the
bias voltage $V$ for different values of the chemical potential.
Bottom figure: matrix elements of the Green's functions:
$G11=\gamma_L\vert (G_{dd}^+)_{11}$; $G12L=\gamma_L\vert (G_{dd}^+)_{12}\vert^2$;
$G12R=\gamma_R\vert (G_{dd}^+)_{12}\vert^2$; $G22R=\gamma_R\vert (G_{dd}^+)_{22}\vert^2$.
The parameters are those introduced in Fig. \ref{fig:hamiltonian}.
}
\end{figure}

The average number of electrons, per spin, at the site $j$ of the device is given by
\begin{equation}
 n_j=\sum_{\lambda=L,R}\int_{-\infty}^\infty dE
(G^+_{dd}\Gamma_\lambda G^-_{dd})_{jj}f(E,\mu_\lambda,T_\lambda)\,.
\end{equation}
Specializing to the sites $j=a'$ and $j=c'$ we obtain
\begin{eqnarray}
n_{a'}= \int_{-\infty}^\infty dE
[\gamma_L\vert (G_{dd}^+)_{11}\vert^2f_L+\gamma_R\vert (G_{dd}^+)_{12}\vert^2f_R]\,,\\
n_{c'}=\int_{-\infty}^\infty dE
[\gamma_L\vert (G_{dd}^+)_{12}\vert^2f_L+\gamma_R\vert (G_{dd}^+)_{22}\vert^2f_R]\,,
\label{NaNc}
\end{eqnarray}
where the $f_{L/R}$ are the Fermi functions, and $\gamma_{L/R}$ are defined as
\begin{eqnarray}
 \gamma_{L}=-\frac{V'^2}{\pi}\Im(G_{LL}^+)_{22}\,,\\
 \gamma_{R}=-\frac{t'^2}{\pi}\Im(G_{RR}^+)_{11}\,.
\end{eqnarray}

From Figure \ref{fig:density} we see that the local electronic density at 
$a'$ and $c'$ sites is a highly non-linear function of the bias potential.
This behavior can only be understood by looking at the
behavior of the matrix elements of the junction Green's function
both as a function of energy and bias voltage. Clearly there are many resonances
in the Green's function matrix elements, which renders the analysis rather
difficult. For small and intermediate bias voltage we have:
\begin{equation}
 \gamma_L\vert (G_{dd}^+)_{11}\gg\gamma_R\vert (G_{dd}^+)_{12}\vert^2\,,
\end{equation}
and
\begin{equation}
 \gamma_L\vert (G_{dd}^+)_{12}\vert^2\gg \gamma_R\vert (G_{dd}^+)_{22}\vert^2\,.
\end{equation}
These inequalities are due to the relatively large value of $\epsilon'_c$
relatively to $\epsilon_c$. On the other hand, when the voltage increases,
the above inequalities transform, roughly, into approximated equalities.

As a general trend, $n_{a'}$ and $n_{c'}$ will decrease at relatively large values of $eV$, due to
the dependence of the energies $\epsilon_{a'}$  and $\epsilon_{c'}$ on $V$
(see Eq. (\ref{EV})); this is specially the case for small coupling between
the semiconductor and the metal (small $W$). In the case of small coupling, the contribution
due to the off-diagonal Green's function is small, since this latter quantity
is proportional to $W$ and therefore its contribution to $n_{a'}$ and $n_{c'}$
(see Eq. (\ref{NaNc})) is proportional to $W^2$. So, in this case, the occupancy of the
sites $a'$ and $c'$ is only due to the electronic wave-function coming from the system to which
the corresponding site is directly connected. 
 
For large bias voltage the matrix elements that contribute
to $n_{c'}$ develop large resonances which contribute to the increase of
$n_{a'}$ relatively to $n_{c'}$ at $V\simeq 3$ eV.

It turns out that the details of the behavior of $n_{a'}$ and $n_{c'}$ depend
some what on the relative strength of the hopping parameters and on-site
energies.
\section{Summary and conclusions}
\label{conclusions}
In this paper we have studied a simple one-dimensional model of semiconductor-metal junction.
The advantage of this simplification is that all the features can be studied using analytical 
expressions. We have shown that the semi-infinite metal does generate surface states in particular
conditions. For the semiconductor surface states can form in the gap.
Resonances  can be formed local density of states of the metal if the surface parameters are
very  different from those in the bulk. The energy position of such resonance is given by
\begin{equation}
E_R=\frac {2t^2\epsilon_{c'}-t'^2\epsilon_{c}}{2t^2-t'^2}\,. 
\end{equation}
In the case $t'=t$ and $\epsilon_{c'}=\epsilon_{c}$, $E_R$ is simply the energy of the
maximum of the local density of states.

When the interface
is formed by making the parameter $W$ finite, the surface state formed in the
gap of the semiconductor becomes a resonance,
because the density of states of the metal is finite in the gap of the semiconductor.
This is a consequence of the choice of the parameters for the semiconductor
for the metal. We could as well have chosen a different set of parameters, such that the
density of states of metal was zero in the gap of the semiconductor. In this case the
surface state, as pole of the full Green's function, would still survive as long as
the renormalization of its energy due to the finiteness of the real part of the metal
Green's function would not move it away from the gap. As $W$ changes  the
resonances in the semiconductor and in the metal drift in energy

These resonances have a strong impact on the transport properties of the junction,
because for fixed values of $W$, their energy location is dependent on the bias
voltage which changes the on-site energies of the atoms in the junction. Their
effect is even more dramatic on the electronic occupancy of the atoms in the
junction.

Although our system is a very simple one, the features seen in this case will also be
present on more realistic cases. However, in the context of quasi one-dimensional organic 
conductors our calculations have direct relevance; for example, the electronic transport
in a molecular wire will develop features as those
described here close to the contacts to the metallic leads.

\section*{Acknowledgments}
This work was supported by FCT under the grant
\newline
 PTDC/FIS/64404/2006.  The author acknowledges Ricardo M.
Ribeiro for a critical reading of the manuscript.

\appendix
\section{Simple derivation of Eq. (\ref{GRR})}

The Hamiltonian of the semi-infinite lead has the form
\begin{equation}
 H=\left[
\begin{array}{ccccc}
H_R & C_R &    &   &\\
C^\dag_R & H_R & C_R & &\\
 & C^\dag_R & H_R &  C_R &\\
 & & &  \ddots &
\end{array}
\right]=
\left[
\begin{array}{cc}
H_R & V_R \\
V^\dag_R& H
 \end{array}
\right]
\end{equation}
Let us define the Green's function of the system by $(E-H)G=\bm 1$.
The $G_{RR}$ Green's function is obtained from
\begin{equation}
\left[
 \begin{array}{cc}
E-H_R & -V_R \\
-V^\dag_R& E-H
 \end{array}
\right]
\left[
 \begin{array}{cc}
G_{RR} & G_{R1} \\
G_{1R}& G_{11}
 \end{array}
\right]=\bm 1\,.
\label{Eq1}
\end{equation}
From Eq. (\ref{Eq1}) we derive
\begin{eqnarray}
 (E-H_R)G_{RR}-V_RG_{1R}=1\,,\\
-V^\dag_RG_{RR}+(E-H)G_{1R}=0\,.
\end{eqnarray}
Solving this linear system for $G_{RR}$ we obtain
\begin{equation}
(E-H_R)G_{RR}-V_R(E-H)^{-1}V^\dag_RG_{RR}=\bm 1\,, 
\end{equation}
but $(E-H)^{-1}=G$ and $V_RGV^\dag_R=C_RG_{RR}C^\dag_R$, which leads to Eq. (\ref{GRR}).


\end{document}